\documentclass[twocolumn,epjc3]{svjour3}
\usepackage{amsfonts}
\usepackage{amsmath}
\usepackage{amssymb,bbold}
\usepackage[normalem]{ulem}

\RequirePackage[T1]{fontenc}
\smartqed
\RequirePackage{graphicx}
\RequirePackage{mathptmx}
\RequirePackage{flushend}
\RequirePackage[numbers,sort&compress]{natbib}
\RequirePackage[colorlinks,citecolor=blue,urlcolor=blue,linkcolor=blue]{hyperref}
\journalname{Eur. Phys. J. C}
\RequirePackage{color}

\begin{document}

\title{Relativistic quantum dynamics of neutral particle in
external electric fields: An approach on effects of spin}

\author{F. S. Azevedo \thanksref{e1,addr1} \and Edilberto O. Silva %
\thanksref{e2,addr2} \and Luis B. Castro \thanksref{e3,addr2} \and Cleverson
Filgueiras\thanksref{e4,addr1,addr3} \and D. Cogollo \thanksref{e5,addr1}}
\thankstext{e1}{e-mail: frankitaly14@gmail.com}
\thankstext{e2}{e-mail:
edilbertoos@pq.cnpq.br} \thankstext{e3}{e-mail: lrb.castro@ufma.br} %
\thankstext{e4}{e-mail: cleversonfilgueiras@yahoo.com.br} %
\thankstext{e5}{e-mail: diegocogollo@df.ufcg.edu.br}

\institute{Departamento de F\'{\i}sica,
           Universidade Federal de Campina Grande,
           Caixa Postal 10071,
           58109-970, Campina Grande, Para\'{i}ba, Brazil \label{addr1}
           \and
           Departamento de F\'{i}sica,
           Universidade Federal do Maranh\~{a}o,
           Campus Universit\'{a}rio do Bacanga,
           65080-805, S\~{a}o Lu\'{i}s, Maranh\~{a}o, Brazil \label{addr2}
           \and
           Departamento de Física (DFI), Universidade Federal de Lavras (UFLA), Caixa Postal 3037, 37200-000, Lavras, Minas Gerais, Brazil \label{addr3}
           }
\date{Received: date / Accepted: date}
\maketitle

\begin{abstract}
The planar quantum dynamics of spin-1/2 neutral particle interacting with
electrical fields is considered. A set of first order differential equations
are obtained directly from the planar Dirac equation with nonminimum
coupling. New solutions of this system, in particular, for the
Aharonov-Casher effect, are found and discussed in detail. Pauli equation is
also obtained by studying the motion of the particle when it describes a
circular path of constant radius. We also analyze the planar dynamics in the
full space, including the $r=0$ region. The self-adjoint extension method is
used to obtain the energy levels and wave functions of the particle for two
particular values for the self-adjoint extension parameter. The energy
levels obtained are analogous to the Landau levels and explicitly depend on
the spin projection parameter.
\end{abstract}

\section{Introduction}

Topological effects in quantum mechanics has been one of the most studied
problems of planar dynamics in recent years. These phenomena present no
classical counterparts and are associated with physical systems defined on a
multiply connected space-time \cite{PRL.1995.74.2847}. Many of the recent
interest in this matter is a consequence of the pioneering work by Aharonov
and Bohm (AB) \cite{PR.1959.115.485}, where they proposed the first example
of generation of topological phase acquired by an electron when it travels
through a magnetic field-free region. This phenomenon, known as
Aharonov-Bohm effect, has been the usual framework for studying properties
of other physical systems which lead to similar effects. The first
subsequent effect was the work by Aharonov-Casher (AC) \cite{PRL.1984.53.319}%
, where they predicted that the wave function of a neutral particle with a
magnetic dipole moment acquires a topological phase when traveling in a
closed path which encircles an infinitely long filament carrying an uniform
charge density. Subsequently, several other AB-like effects were discovered
over the last three decades (see Refs. \cite%
{PRL.1987.58.1593,PRL.1990.65.1697,PRA.1993.47.3424,PRL.1995.75.2071,PRL.1994.72.5, PRL.2000.85.1354}
).

An important question that we address here are the effects of spin on the
dynamics of topological effects. The first work in this context was proposed
by Hagen to study the scattering of relativistic spin-1/2 particles in an AB
potential \cite{PRL.1990.64.503}. He showed that, by reformulating the
problem with a source of finite radius which is then allowed to go to zero,
it is established that the delta function alone that causes solutions that
are singular at the origin. He also concluded that the modifications in the
amplitude which arise from the inclusion of spin are seen to modify the
cross section for the case of polarized beams. Hagen has also shown that
there is an exact equivalence between the AB effect for spin-1/2 particles
and the AC effect \cite{PRL.1990.64.2347}. This fact establishes the
dynamics of the AC problem. However, a peculiarity that Hagen has not
addressed clearly in their work is how to find the bound states energy
levels. By modeling the problem by boundary conditions at the origin, an
expression for the bound state energy for the AC problem was derived in Ref.
\cite{EPJC.2013.73.2402}. The method used to find these energies was
established in Refs. \cite{PRD.2012.85.041701,AoP.2013.339.510}, and it is
based on the self-adjoint extension method of operators in quantum mechanics.

In the AC problem, the electric field is the one generated by an infinitely
long, infinitesimally thin line of charge along the $z$-axis with a charge
density $\lambda _{1}$ distributed uniformly about it, namely
\begin{equation}
\mathbf{E}_{1}=2\lambda _{1}\frac{\mathbf{\hat{\rho}}}{\rho },~~~\mathbf{%
\nabla }\cdot \mathbf{E}_{1}=2\lambda _{1}\frac{\delta (\rho )}{\rho },
\label{fE}
\end{equation}%
As pointed out in Ref. \cite{PRL.1990.64.2347}, to ensure the exact
equivalence between the AB and AC effects, we can not neglect the $%
\boldsymbol{\nabla }\cdot \mathbf{E}_{1}$ term. The physical implications of
this term on the dynamics of the particle has been quite studied in recent
years \cite%
{EPL.2013.101.51005,EPJC.2014.74.2708,JPA.2010.43.354008,JPG.2013.40.075007}.

Another configuration field of interest is%
\begin{equation}
\mathbf{E}_{2}=\frac{\lambda _{2}\rho }{2}\mathbf{\hat{\rho}},~~~\mathbf{%
\nabla }\cdot \mathbf{E}_{2}=\lambda _{2},~\frac{\partial \mathbf{E}_{2}}{%
\partial t}=0,~~~\mathbf{\nabla }\times \mathbf{E}_{2}=0.  \label{fEb}
\end{equation}%
This special configuration was proposed by Ericsson and Sj\"{o}qvist to
study an atomic analog of the Landau quantization based on the AC effect
\cite{PRA.2001.65.013607}. They demonstrated that the existence of a certain
field-dipole configuration in which an atomic analog of the standard Landau
effect occurs opens up the possibility for an atomic realization of the
quantum Hall effect using electric fields. This same configuration was used
to study the Landau levels in the nonrelativistic dynamics of a neutral
particle which possesses a permanent magnetic dipole moment interacting with
an external electric field in the curved spacetime background with either
the presence or the absence of a torsion field \cite{PRD.2009.79.024008}
(see also Ref. \cite{PRA.2009.80.032106}).

In this article, we analyze the planar motion of a neutral particle of
spin-1/2 interacting with both\textbf{\ }electric fields of Eqs. (\ref{fE})
and (\ref{fEb}), i.e.,%
\begin{equation}
\mathbf{E}=\mathbf{E}_{1}+\mathbf{E}_{2}.  \label{cE}
\end{equation}%
Although the field configuration (\ref{fE}) has been studied in different
contexts in the literature, in our approach, we solve the first order Dirac
equation and derive their solutions giving a focus to the effects due to the
spin. This analysis, in particular to the electric field of Eq. (\ref{fE}),
which is responsible for the AC problem, it is presented and discussed in
detail here for the first time. We also address the second-order Dirac
equation, which results in the Pauli equation. We make use of the
self-adjoint extension method \cite{Book.1975.Reed.II} and model the
Hamiltonian by boundary conditions \cite{CMP.1991.139.103}. We also
determine an expression for the energy levels of the particle and compare it
with the known results in the literature.

This paper is organized as follows. In Sec. \ref{motion}, we consider the
Dirac equation with nonminimal coupling and construct the set of first order
differential equations. In Sec. \ref{isolated}, we solve the first order
differential equations and obtain the bound state solutions of the particle.
The existence of these solutions means that the system admits isolated
solutions. In Sec. \ref{sqdirac}, we derive the second-order equation (Pauli
equation) and solve it by assuming that the particle describes a circular
path of constant radius. In Sec. \ref{selfae}, we analyze the dynamics of
the system with the inclusion of the $r=0$ region. We use the self-adjoint
extension method to fix the physics of the problem in the $r=0$ region.
Expressions for the wave functions and energies are obtained, without any
arbitrary parameter which arises from the self-adjoint extension approach.
In Sec. \ref{nonrel}, the results of Sec. \ref{selfae} are examined in the
nonrelativistic limit. In Sec. \ref{conc}, we present our conclusions.

\section{Equation of motion}

\label{motion}

We start with the Dirac equation with nonminimal coupling ($\hbar =c=1$)%
\begin{equation}
\left[ i\gamma ^{\mu }\partial _{\mu }-\frac{\mu }{2}\sigma ^{\mu \nu
}F_{\mu \nu }-M\right] \psi =0,  \label{d1}
\end{equation}%
where $\mu $ is the magnetic dipole moment of the particle, $F_{\mu \nu }$
is the electromagnetic tensor whose components are given by%
\begin{eqnarray}
F_{0i} &=&E^{i}=-E_{i},  \label{compE} \\
F_{ij} &=&-\varepsilon _{ijk}B^{k}=\varepsilon _{ijk}B_{k},  \label{compB}
\end{eqnarray}%
and%
\begin{equation}
\left( \sigma ^{0j},\sigma ^{ij}\right) =\left( i\alpha ^{j},-\epsilon
_{ijk}\Sigma ^{k}\right) ,
\end{equation}%
where $\Sigma ^{k}$ is the spin vector, are the components of the operator%
\begin{equation}
\sigma ^{\mu \nu }=\frac{i}{2}\left[ \gamma ^{\mu },\gamma ^{\nu }\right] .
\end{equation}%
As the particle interacts only with electric fields, we consider only (\ref%
{compE}). In the above representation, Eq. (\ref{d1}) can be written as%
\begin{equation}
\left[ \beta \gamma ^{i}p_{i}+\beta m-i\mu \beta \gamma ^{i}E_{i}\right]
\psi =\mathcal{E}\psi .  \label{d2}
\end{equation}%
Following Ref. \cite{PRL.1990.64.503}, we write $\gamma ^{i}$ as
\begin{equation}
\gamma ^{i}=is\epsilon _{ij}\beta \gamma ^{j},
\end{equation}%
so that Eq. (\ref{d2}) becomes
\begin{equation*}
\left[ \beta \gamma ^{i}p_{i}+\beta m-\mu s\beta \gamma ^{i}\check{E}_{i}%
\right] \psi =\mathcal{E}\psi ,
\end{equation*}%
where $\check{E}_{i}=\epsilon _{ij}E_{j}$, $\epsilon _{ij}=-\epsilon _{ji}$.
Equation (\ref{d2}) can be written as
\begin{equation}
\left[ \beta \boldsymbol{\gamma }\cdot \left( \mathbf{p}-\mu s\boldsymbol{%
\check{E}}\right) +\beta m\right] \Psi =\mathcal{E}\psi .  \label{d3}
\end{equation}%
The $\gamma $ matrices are conveniently defined in terms of the Pauli
matrices \ as \cite{PRL.1990.64.503}
\begin{equation}
\beta \gamma _{1}=\sigma _{1},\qquad \beta \gamma _{2}=s\sigma _{2},\qquad
\beta =\sigma _{3},  \label{matrices}
\end{equation}%
where $s$ is twice the spin value, with $s=+1$ for spin \textquotedblleft
up\textquotedblright\ and $s=-1$ for spin \textquotedblleft
down\textquotedblright . Thus, Eq. (\ref{d3}) can be written as
\begin{equation}
\left[ \sigma _{1}\left( p_{1}-\mu s\check{E}_{1}\right) +s\sigma _{2}\left(
p_{2}-\mu s\check{E}_{2}\right) +\sigma _{3}M\right] \psi =\mathcal{E}\psi .
\label{d4}
\end{equation}%
By noting that $\check{E}_{1}=E_{2}$ e $\check{E}_{2}=-E_{1}$, as usual, we
write Eq. (\ref{d4}) in polar coordinates $(\rho ,\varphi )$
\begin{align}
& \left[ -i\frac{\partial }{\partial \rho }-is\frac{1}{\rho }\frac{\partial
}{i\partial \varphi }-i\left( \frac{\eta _{1}}{\rho }+\eta _{2}\rho \right) %
\right] \psi _{2}  \notag \\
& =\mathrm{e}^{+is\phi }\left( \mathcal{E}-M\right) \psi _{1},  \label{e3} \\
& \left[ -i\frac{\partial }{\partial \rho }+is\frac{1}{\rho }\frac{\partial
}{i\partial \varphi }+i\left( \frac{\eta _{1}}{\rho }+\eta _{2}\rho \right) %
\right] \psi _{1}  \notag \\
& =\mathrm{e}^{-is\phi }\left( \mathcal{E}+M\right) \psi _{2},  \label{e4}
\end{align}%
where $\eta _{1}=2\mu \lambda _{1},~\eta _{2}=\mu \lambda _{2}/2$. Using the
decomposition
\begin{equation}
\psi =\left(
\begin{array}{c}
\psi _{1} \\
\psi _{2}%
\end{array}%
\right) =\left[
\begin{array}{c}
\sum_{m}f_{m}(\rho )e^{im\phi } \\
\sum_{m}ig_{m}(\rho )e^{i(m+s)\phi }%
\end{array}%
\right] ,  \label{ansatz}
\end{equation}%
where $m=0,\pm 1,\pm 2,\pm 3,\ldots $ is the angular momentum quantum
number, Eqs. (\ref{e3})-(\ref{e4}) provide two coupled first-order radial
equations
\begin{align}
& \left[ \frac{d}{d\rho }+\frac{sm+\eta _{1}+1}{\rho }+\eta _{2}\rho \right]
g_{m}(\rho )=\left( \mathcal{E}-M\right) f_{m}(\rho ),  \label{e5} \\
& \left[ -\frac{d}{d\rho }+\frac{sm+\eta _{1}}{\rho }+\eta _{2}\rho \right]
f_{m}(\rho )=\left( \mathcal{E}+M\right) g_{m}(\rho ).  \label{e6}
\end{align}%
The factor $i$ on the lower spinor component in Eq. (\ref{ansatz}) was
inserted to ensure that the radial part of the spinors is manifestly real.
An isolated solution for the problem can be obtained considering the
particle at rest, i.e., $\mathcal{E}=\pm M$. Such solution for the Dirac
equation in $(1+1)$ dimensions was investigated in Ref. \cite%
{AoP.2013.338.278} (see also Refs. \cite{JPA.2007.40.263,PLA.2006.351.379}).

\section{Isolated solutions and the Aharonov-Casher problem}

\label{isolated}

In order to obtain isolated solutions, let us look for bound state solutions
subjected to the normalization condition
\begin{equation}
\int_{0}^{\infty }\left( |f_{m}(\rho )|^{2}+|g_{m}(\rho )|^{2}\right) \rho
d\rho =1,  \label{norm}
\end{equation}%
and consider the conditions $\mathcal{E}=\pm M$ stated above.

\subsection{Case $\mathcal{E}=M$}

In this case, Eqs. (\ref{e5})-(\ref{e6}) are written as
\begin{eqnarray}
&&\left[ \frac{d}{d\rho }+\frac{sm+\eta _{1}+1}{\rho }+\eta _{2}\rho \right]
g_{m}(\rho )=0,  \label{eq1} \\
&&\left[ -\frac{d}{d\rho }+\frac{sm+\eta _{1}}{\rho }+\eta _{2}\rho \right]
f_{m}(\rho )=2Mg_{m}(\rho ).  \label{eq2}
\end{eqnarray}%
The solutions to $g_{m}(\rho )$ and $f_{m}(\rho )$ are
\begin{align}
& g_{m}(\rho )=c_{2}\,\rho ^{-\left( ms+\eta _{1}+{1}\right) }\,e^{-\frac{%
\eta _{2}}{2}\rho ^{2}}{,}  \label{soli1} \\
& f_{m}(\rho )=\rho ^{ms+\eta _{1}}~e^{\frac{\eta _{2}}{2}\rho ^{2}}  \notag
\\
& \times \left[ c_{1}+c_{2}M\left( \eta _{2}\right) ^{ms+\eta _{1}}~\Gamma
\left( -sm-\eta _{1},\eta _{2}\rho ^{2}\right) \right] ,  \label{soli2}
\end{align}%
where $c_{1}$ and $c_{2}$ are constants, and $\Gamma \left( -sm-\eta
_{1},\eta _{2}\rho ^{2}\right) $ is the upper incomplete Gamma function,
obtained through the relation \cite{Book.1972.Abramowitz}
\begin{equation}
\Gamma (a,x)=\int_{x}^{\infty }t^{a-1}e^{-t}dt,\qquad \Re (a)>0.
\label{eq:incgamma}
\end{equation}%
As $\eta _{1,2}\gtrless 0$, then $g_{m}(\rho )$ converges as $\rho
\rightarrow 0$. Moreover, since $\Gamma \left( -sm-\eta _{1},\eta _{2}\rho
^{2}\right) $ always diverges, then $f_{m}(\rho )$ will only converge if $%
c_{2}=0$ e $\eta _{2}<0$. As a result, we have
\begin{equation}
\left[
\begin{array}{c}
f_{m}(\rho ) \\
g_{m}(\rho )%
\end{array}%
\right] =c_{1}\left(
\begin{array}{c}
1 \\
0%
\end{array}%
\right) \rho ^{\eta _{1}+ms}e^{\frac{\eta _{2}}{2}\rho ^{2}},~\left\{
\begin{tabular}{l}
\hspace{-0.3cm} $s=\pm 1,$ \\
\hspace{-0.3cm} $\eta _{1}\gtrless 0,$ \\
\hspace{-0.3cm} $\eta _{2}<0,$ \\
\hspace{-0.3cm} $c_{2}=0,$%
\end{tabular}%
\ \right. .  \label{iso1}
\end{equation}

\subsection{Case $\mathcal{E}=-M$}

In this case, Eqs. (\ref{e5})-(\ref{e6}) become
\begin{align}
\left[ \frac{d}{d\rho }+\frac{sm+\eta _{1}+1}{\rho }+\eta _{2}\rho \right]
g_{m}(\rho )& =-2Mf_{m}(\rho ),  \label{eq3} \\
\left[ -\frac{d}{d\rho }+\frac{sm+\eta _{1}}{\rho }+\eta _{2}\rho \right]
f_{m}(\rho )& =0.  \label{eq4}
\end{align}%
The solutions of these equation are%
\begin{align}
& f_{m}\left( \rho \right) =c_{1}\rho ^{ms+\eta _{1}}~e^{\frac{\eta _{2}}{2}%
\rho ^{2}},  \label{soli3} \\
& g_{m}(\rho )=e^{-\frac{1}{2}\eta _{2}\rho ^{2}}\rho ^{-\left( sm+\eta
_{1}+1\right) }  \notag \\
& \times \left[ c_{2}+c_{1}M\left( -\eta _{2}\right) {}^{-\left( sm+\eta
_{1}+1\right) }~\Gamma \left[ ms+\eta _{1}+1,-\eta _{2}\rho ^{2}\right] %
\right] .  \label{soli4}
\end{align}%
As for the case $\mathcal{E}=M$, looking for solutions ( \ref{soli3})\ and (%
\ref{soli4}), we can see that the only square integrable solutions are
\begin{equation}  \label{iso2}
\left[
\begin{array}{c}
f_{m}(\rho ) \\
g_{m}(\rho )%
\end{array}%
\right] =c_{2}\left(
\begin{array}{c}
0 \\
1%
\end{array}%
\right) e^{-\frac{1}{2}\eta _{2}\rho ^{2}}\rho ^{-\left( sm+\eta
_{1}+1\right) },~\left\{
\begin{tabular}{l}
\hspace{-0.3cm} $s=\pm 1,$ \\
\hspace{-0.3cm} $\eta _{1}\gtrless 0,$ \\
\hspace{-0.3cm} $\eta _{2}>0,$ \\
\hspace{-0.3cm} $c_{1}=0,$%
\end{tabular}%
\right. .
\end{equation}
In summary, note that the above results, Eqs. (\ref{iso1}) and (\ref{iso2})
are bound-state solutions of square-integrable because the function $e^{%
\frac{\pm\eta_{2}}{2}\rho^{2}}$ with $\eta_{2}\lessgtr 0$ predominates over
the polynomials $\rho^{\eta_{1}+ms}$ and $\rho^{-(sm+\eta_{1}+1)}$ for $%
\mathcal{E}=M$ and $\mathcal{E}=-M$, respectively. We can conclude that the
presence of the $\lambda_{2}$ is necessary for the existence of bound states.

\section{The quadratic equation}

\label{sqdirac}

The equation of second order derivative of Eq. (\ref{d3}) is found to be%
\begin{equation}
\left[ \mathbf{p}+s(\mathbf{\mu }\times \mathbf{E})\right] ^{2}\psi +\mu
\sigma _{3}(\mathbf{\nabla }\cdot \mathbf{E})\psi =\left( \mathcal{E}%
^{2}-{}M^{2}\right) \psi .  \label{pauliequation}
\end{equation}%
Using (\ref{cE}), Eq. (\ref{pauliequation}) can be written more explicitly
as
\begin{eqnarray}
&&\Bigg[-\frac{\partial ^{2}}{\partial \rho ^{2}}-\frac{1}{\rho }\frac{%
\partial }{\partial \rho }-\frac{1}{\rho ^{2}}\frac{\partial ^{2}}{\partial
\varphi ^{2}}+\,2s\left( \frac{\eta _{1}}{\rho }+\eta _{2}\rho \right) \frac{%
1}{\rho }\frac{\partial }{i\partial \varphi }  \notag \\
&&+\left( \frac{\eta _{1}}{\rho }+\eta _{2}\rho \right) ^{2}\Bigg]\psi
+\sigma _{z}\left[ \eta _{1}\frac{\delta \left( \rho \right) }{\rho }+2\eta
_{2}\right] \psi  \notag \\
&=&\left( \mathcal{E}^{2}-{}M^{2}\right) \psi ,  \label{pauli}
\end{eqnarray}%
In this stage, it is worthwhile to mention that Eq. (\ref{pauli}) is the
correct quadratic form of the Dirac equation with nonminimal coupling,
because the singular term from $\mathbf{\nabla }\cdot \mathbf{E}_{1}$ is
considered.

Using the decomposition (\ref{ansatz}), the equation for $f_{m}(\rho )$ can
be obtained
\begin{equation}
hf_{m}(\rho )=Ef_{m}(\rho ),  \label{eqpauli}
\end{equation}%
with $E=\mathcal{E}^{2}-M^{2}-2\eta _{2}\left( \eta _{1}+ms+1\right) $,
where $\ $%
\begin{equation}
h=h_{0}+\eta _{1}\frac{\delta \left( \rho \right) }{\rho },  \label{hpauli}
\end{equation}%
is the Hamiltonian system with the magnetic moment of the particle pointing
the positive direction of z axis, and
\begin{equation}
h_{0}=-\frac{d^{2}}{d\rho ^{2}}-\frac{1}{\rho }\frac{d}{d\rho }+\frac{\nu
^{2}}{\rho ^{2}}+\eta _{2}^{2}\rho ^{2}.  \label{hzpauli}
\end{equation}%
is Hamiltonian without the $\delta $ function, and
\begin{equation}
\nu =m+s\eta _{1}.
\end{equation}%
The equation for $g_{m}(\rho )$ is obtained in an immediate way. It is given
by%
\begin{equation}
\bar{h}g_{m}(\rho )=\bar{E}g_{m}(\rho ),  \label{pauli2}
\end{equation}%
with $\bar{E}=\left( \mathcal{E}^{2}-M^{2}\right) -2\eta _{2}\left[ s\left(
m+s\right) +\eta _{1}-1\right] $, where $\ $%
\begin{equation}
\bar{h}=\bar{h}_{0}-\eta _{1}\frac{\delta \left( \rho \right) }{\rho },
\label{hpauli2}
\end{equation}%
\begin{equation}
\bar{h}_{0}=-\frac{d^{2}}{d\rho ^{2}}-\frac{1}{\rho }\frac{d}{d\rho }+\frac{%
\bar{\nu}^{2}}{\rho ^{2}}+\eta _{2}^{2}\rho ^{2},  \label{hzero2}
\end{equation}%
and
\begin{equation}
\bar{\nu}=m+s+s\eta _{1}.  \label{angulard}
\end{equation}%
Equation (\ref{pauli}) governs the system dynamics. In this dynamic, we must
consider regular and irregular solutions, since irregular solutions are also
physical solutions for the system under consideration. In other words, since
we consider the effects of the spin of the particle and because of the field
configuration (\ref{fE}), the Hamiltonian will contain a singular potential.
We will return to this problem in Section \ref{selfae}.

\subsection{Particle in a ring of constant radius}

\begin{figure}[tbp]
\includegraphics[width=0.73\linewidth,angle=270]{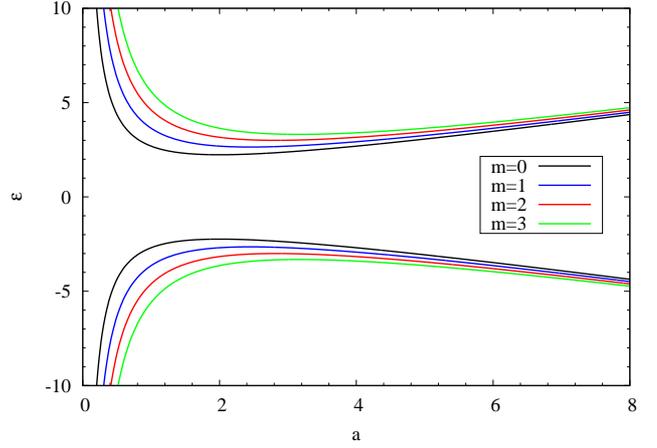}
\caption{Plots of the energy as a function of radius for $\protect\varsigma %
=1$ and different values of $m$.}
\label{fig1}
\end{figure}

Before solving Eq. (\ref{pauli}), an interesting case which can be
considered here is when we assume the particle describing a circular motion
of radius $\rho =a=const.$. In this case, from (\ref{pauli}) we get%
\begin{equation}
\left[ \frac{1}{a}\frac{\partial }{i\partial \varphi }+s\left( \frac{\eta
_{1}}{a}+\eta _{1}a\right) \right] ^{2}\psi =\left( \mathcal{E}%
^{2}-{}M^{2}\right) \psi .
\end{equation}%
According to Eq. (\ref{ansatz}), the wave function for a particle executing
a circular motion of constant radius can be written as%
\begin{equation}
\psi =\left[
\begin{array}{c}
a_{m} \\
ib_{m}e^{is\phi }%
\end{array}%
\right] e^{im\phi }.
\end{equation}%
If $\psi $ is an eigenvector of $\sigma _{z}$ with eigenvalue $\varsigma
=\pm 1$, the energy levels are given by
\begin{align}
\mathcal{E}^{2}-{}M^{2}& =\left[ \frac{m}{a}+s\left( \frac{\eta _{1}}{a}%
+\eta _{2}a\right) \right] ^{2},~~\left( \varsigma =+1\right)
\label{espinup} \\
\mathcal{E}^{2}-{}M^{2}& =\left[ \frac{m+s}{a}+s\left( \frac{\eta _{1}}{a}%
+\eta _{2}a\right) \right] ^{2},~~\left( \varsigma =-1\right)
\label{espindown}
\end{align}%
The profiles of the energy is shown in Fig. \ref{fig1} for $\varsigma =1$
and some values of $m$. Figure \ref{fig1} clearly shows that both particle
and antiparticle energy levels are members of the spectrum. Note that for
positive (negative)-energy we find that the lowest quantum number
corresponds to the lowest (highest) energies, so that it is plausible to
identify them with particle (antiparticle) energy levels. Also, it is
noticeable that the Dirac energies are symmetrical about $\mathcal{E}=0$ and
since the positive and negative energies never intercept one can see that
there is no channel for spontaneous particle-antiparticle creation.
If $\eta _{2}=0$, we obtain the energy levels of a neutral particle with
magnetic moment $\mu $ in a circular path of constant radius,
\begin{eqnarray}
\mathcal{E}^{2}-{}M^{2} &=&\frac{1}{a^{2}}\left( m+s\eta _{1}\right) ^{2}, \\
\mathcal{E}^{2}-{}M^{2} &=&\frac{1}{a^{2}}\left( m+s+s\eta _{1}\right) ^{2}.
\end{eqnarray}%
These energies correspond to spectrum for the usual Aharonov-Casher effect.

\section{Self-adjoint extension analysis and the dynamic including the $r=0$
region}

\label{selfae}

In this section, we solve Eqs. (\ref{eqpauli}) and (\ref{pauli2}), including
the term $\delta \left( \rho \right) /\rho $. In order to deal with this
kind of point interaction potential, we consider the self-adjoint extension
approach \cite{Book.2004.Albeverio,JMP.1985.26.2520}. In quantum mechanics,
observables correspond to self-adjoint operators. However, in some physical
systems, we deal with differential operators for which the Hamiltonian is
not necessarily symmetrical in some region of the space. In such cases, the
Hamiltonian is not essentially self-adjoint and one attempts to find
self-adjoint extensions of the Hamiltonian corresponding to different types
of boundary conditions. Such self-adjoint extensions are based in boundary
conditions at the origin and conditions at infinity \cite%
{crll.1987.380.87,JMP.1998.39.47,LMP.1998.43.43}. From the theory of
symmetric operators, it is a well-known fact that the symmetric radial
operator $h_{0}$ (as in Eq. (\ref{hpauli})) is essentially self-adjoint for $%
\left\vert \nu \right\vert \geq 1$, while for $\left\vert \nu \right\vert <1$
it admits an one-parameter family of self-adjoint extensions \cite%
{Book.1975.Reed.II}, $h_{0,\lambda _{m}}$, where $\lambda _{m}$ is the
self-adjoint extension parameter. Here, we will use the approach of Ref.
\cite{JMP.1985.26.2520,Book.2004.Albeverio}, which is based in a boundary
conditions at the origin. Thus, all the self-adjoint extensions $%
h_{0,\lambda _{m}}$ of $h_{0}$ are parametrized by the boundary condition at
the origin
\begin{equation}
\varkappa _{0}=\lambda _{m}\varkappa _{1},  \label{bc}
\end{equation}%
with
\begin{align}
\varkappa _{0}={}& \lim_{\rho \rightarrow 0^{+}}\rho ^{|\nu |}f_{m}(\rho ),
\\
\varkappa _{1}=& \lim_{\rho \rightarrow 0^{+}}\frac{1}{\rho ^{|\nu |}}\left[
f_{m}(\rho )-\varkappa _{0}\frac{1}{\rho ^{|\nu |}}\right] .
\end{align}%
For $\lambda _{m}=0$, we have the free Hamiltonian, i.e., without the $%
\delta $ function, with regular wave functions at the origin; for $\lambda
_{m}\neq 0$, the boundary condition in Eq. (\ref{bc}) permit an $\rho
^{-|\nu |}$ singularity in the wave functions at the origin. Thus, by making
a variable change, $\tilde{\rho}=\eta _{2}\rho ^{2}$, Eq. (\ref{eqpauli})
reads
\begin{equation}
\left[ \tilde{\rho}\frac{d^{2}}{d\tilde{\rho}^{2}}+\frac{d}{d\tilde{\rho}}%
-\left( \frac{\nu ^{2}}{4\tilde{\rho}}+\frac{\tilde{\rho}}{4}-\frac{E}{4\eta
_{2}}\right) \right] f_{m}(\tilde{\rho})=0.  \label{edofrho}
\end{equation}%
As mentioned above, the boundary condition (\ref{bc}) allows us to look for
regular and irregular solutions for Eq. (\ref{edofrho}). By studying the
asymptotic limits of Eq. (\ref{edofrho}), we find the solution
\begin{equation}
f_{m}(\tilde{\rho})=\tilde{\rho}^{\pm \frac{\left\vert \nu \right\vert }{2}%
}e^{-\frac{\tilde{\rho}}{2}}F(\tilde{\rho}),  \label{frho}
\end{equation}%
where ($\pm $) refers to the regular (irregular) solution, respectively.
Substituting Eq. (\ref{frho}) into Eq. (\ref{edofrho}), we find%
\begin{equation}
\tilde{\rho}\frac{d^{2}F}{d{\tilde{\rho}}^{2}}+\left( 1\pm \left\vert \nu
\right\vert -\tilde{\rho}\right) \frac{dF(\tilde{\rho})}{d{\tilde{\rho}}}%
-\left( \frac{1\pm \left\vert \nu \right\vert }{2}-\frac{E}{4\eta _{2}}%
\right) F(\tilde{\rho})=0.  \label{edor}
\end{equation}%
Equation (\ref{edor}) is of the confluent hypergeometric equation type
\begin{equation}
zF^{\prime \prime }(z)+(b-z)F^{\prime }(z)-aF(z)=0.
\end{equation}%
In this manner, the general solution for Eq. (\ref{edofrho}) is given by
\begin{eqnarray}
f_{m}(\tilde{\rho}) &=&a_{m}\tilde{\rho}^{\frac{\left\vert \nu \right\vert }{%
2}}e^{-\frac{\tilde{\rho}}{2}}\;F\left( \frac{1+\left\vert \nu \right\vert }{%
2}-\frac{E}{4\eta _{2}},1+\left\vert \nu \right\vert ,\tilde{\rho}\right)
\notag \\
&+&b_{m}\tilde{\rho}^{-\frac{\left\vert \nu \right\vert }{2}}e^{-\frac{%
\tilde{\rho}}{2}}\;F\left( \frac{1-\left\vert \nu \right\vert }{2}-\frac{E}{%
4\eta _{2}},1-\left\vert \nu \right\vert ,\tilde{\rho}\right) .
\label{solgen}
\end{eqnarray}%
In Eq. (\ref{solgen}), $F(a,b,z)$ is the confluent hypergeometric function
of the first kind \cite{Book.1972.Abramowitz} and $a_{m}$ and $b_{m}$ are,
respectively, the coefficients of the regular and irregular solutions.

Now, we remark that Eq. (\ref{edor}) is equivalent to Eq. (38) of Ref. \cite%
{EPJC.2014.74.3187}, in which the procedure for obtaining the energy levels
for different values of the self-adjoint extension parameter is given in
detail. In this procedure, we use the boundary condition (\ref{bc}) together
with the normalizability condition to obtain a relation that allows us to
eliminate $a_{m}$ and $b_{m}$ of Eq. (\ref{solgen}). Then, following Ref.
\cite{EPJC.2014.74.3187}, such condition is found to be%
\begin{equation}
\frac{\Gamma (\frac{1+\left\vert \nu \right\vert }{2}-\frac{E}{4\eta _{2}})}{%
\Gamma (\frac{1-\left\vert \nu \right\vert }{2}-\frac{E}{4\eta _{2}})}=-%
\frac{1}{\lambda _{m}\left( \eta _{2}\right) ^{\left\vert \nu \right\vert }}%
\frac{\Gamma (1+\left\vert \nu \right\vert )}{\Gamma (1-\left\vert \nu
\right\vert )}.  \label{condfe}
\end{equation}%
Equation (\ref{condfe}) gives a contribution of the irregular solution to
the problem. This feature comes from the fact that the operator $H_{0}$ is
not self-adjoint for $\left\vert \nu \right\vert <1$.

We now analyze the following points in Eq. (\ref{condfe}):

(i) For $\lambda _{m}=0$, case in which the $\delta $ function is absent,
only the regular solution contributes for the bound state wave function.

(ii) For $\lambda _{m}=\infty $, only the irregular solution contributes for
the bound state wave function.

Thus, for all the other values of the self-adjoint extension parameter, both
regular and irregular solutions contributes for the bound state wave
function. Analyzing the poles of the Gamma function in Eq. (\ref{condfe})
together with the criteria (i) and (ii), we get%
\begin{eqnarray}
\frac{1+\left\vert \nu \right\vert }{2}-\frac{E}{4\eta _{2}} &=&-n\text{,}~%
\text{for }\lambda _{m}=0\text{,}  \label{pole1} \\
\frac{1-\left\vert \nu \right\vert }{2}-\frac{E}{4\eta _{2}} &=&-n\text{,
for }\lambda _{m}=\infty \text{,}  \label{pole2}
\end{eqnarray}%
with $n$ a nonnegative integer, $n=0,1,2,\ldots $. By solving Eqs. (\ref%
{pole1}) and (\ref{pole2}) for $\mathcal{E}^{2}-M^{2}$, we obtain,
respectively, for the regular and irregular solutions
\begin{eqnarray}
\frac{\mathcal{E}^{2}-{}M^{2}}{2\eta _{2}} &=&\left( 2n+1+\left\vert m+s\eta
_{1}\right\vert \right) +\eta _{1}+ms+1,  \label{enyf1} \\
\frac{\mathcal{E}^{2}-{}M^{2}}{2\eta _{2}} &=&\left( 2n+1-\left\vert m+s\eta
_{1}\right\vert \right) +\eta _{1}-ms+1.  \label{enyf2}
\end{eqnarray}%
As an illustration, the profiles of energy as a function of $\lambda _{1}$
and with spin projection parameter $s=1$ and $s=-1$ are shown in figures \ref%
{fig2} and \ref{fig3}, respectively. Once again, we note that both particle
and antiparticle energy levels are members of the spectrum. Also, it is
noticeable that in both figures the Dirac energies are symmetrical about $%
\mathcal{E}=0$ and, since the positive and negative energies never
intercept, we can see that there is no channel for spontaneous
particle-antiparticle creation. In this case, from the requirement of real
energies (from equation (\ref{enyf1})) we obtain a constraint on the minimum
value of $\lambda _{1}$. The parameter $\lambda _{1}$ has to satisfies the
following inequation:
\begin{equation}
|m+2\mu s\lambda _{1}|+2\mu \lambda _{1}\geqslant -\left[ \frac{M^{2}}{\mu
\lambda _{2}}+2n+2+sm\right] \,.  \label{vinc}
\end{equation}

\begin{figure}[tbp]
\includegraphics[width=0.73\linewidth,angle=270]{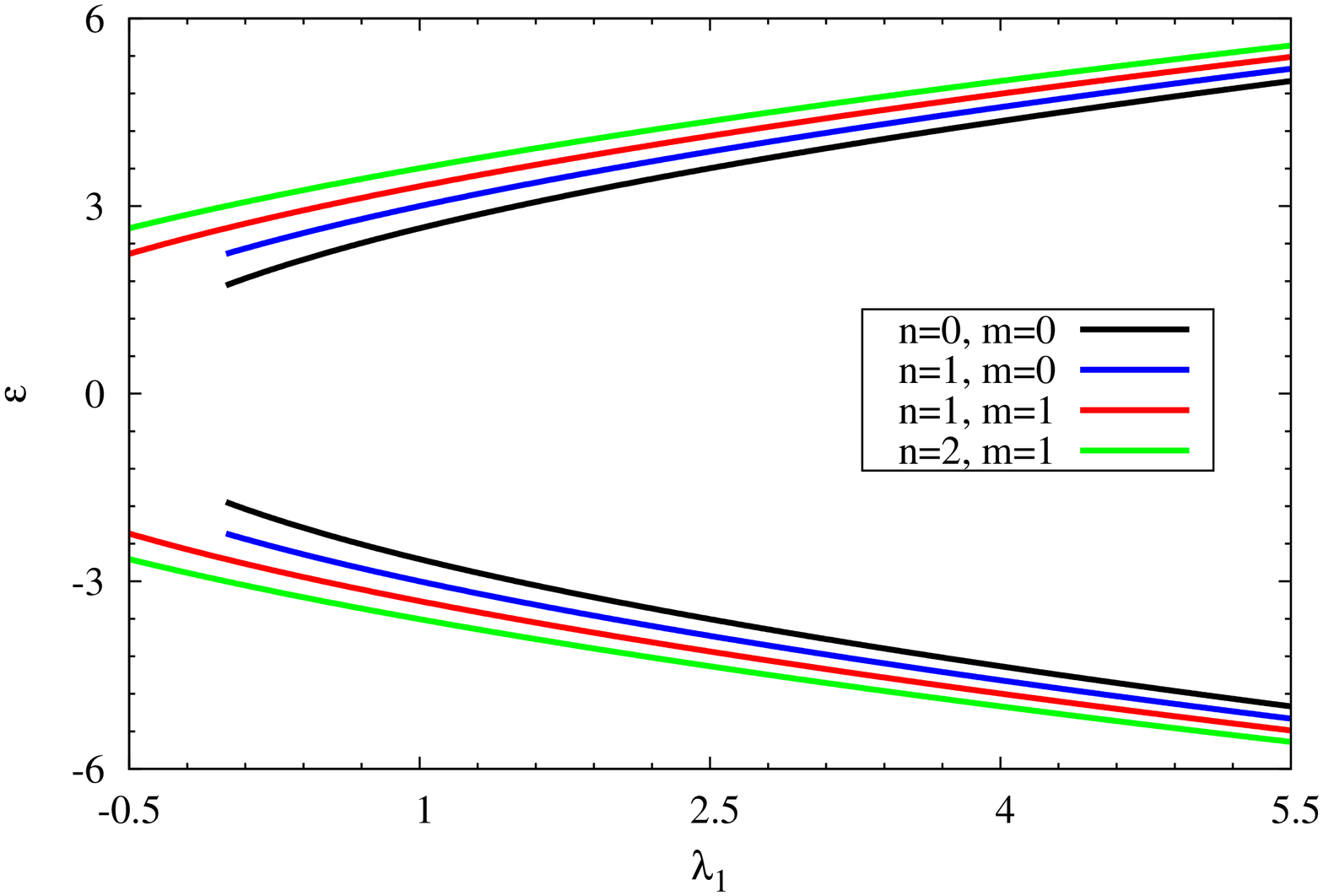}
\caption{Plots of the energy as a function of $\protect\lambda_{1}$ for $s=1$
and different values of $n$ and $m$.}
\label{fig2}
\end{figure}
\begin{figure}[tbp]
\includegraphics[width=0.73\linewidth,angle=270]{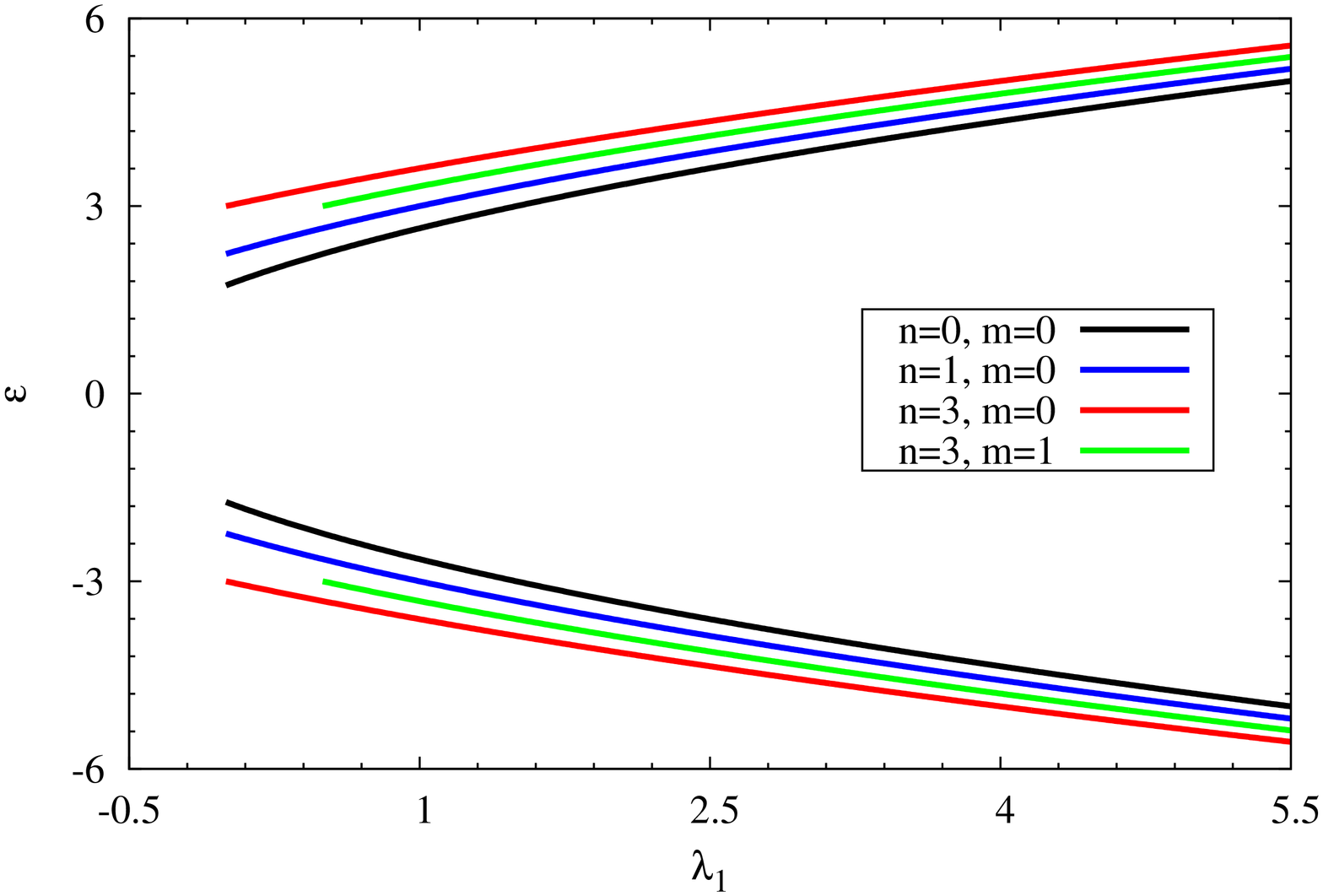}
\caption{Plots of the energy as a function of $\protect\lambda_{1}$ for $%
s=-1 $ and different values of $n$ and $m$.}
\label{fig3}
\end{figure}

The unnormalized bound state wave functions are given by%
\begin{eqnarray}
f_{m}(\tilde{\rho}) &=&\tilde{\rho}^{\frac{\nu }{2}}e^{-\frac{\tilde{\rho}}{2%
}}F\left( -n,1+\nu ,\tilde{\rho}\right)  \notag \\
&+&\tilde{\rho}^{-\frac{\nu }{2}}e^{-\frac{\tilde{\rho}}{2}}F\left( -n,1-\nu
,\tilde{\rho}\right) .
\end{eqnarray}%
Note that when $\left\vert \nu \right\vert \geq 1$ or equivalently when the $%
\delta $ interaction is absent, only the regular solution contributes for
the bound state wave function ($b_{m}=0$), and the energy is given by Eq. (%
\ref{enyf1}).

Now, we consider the solution of Eq. (\ref{pauli2}). By performing the same
steps to achieve Eqs. (\ref{enyf1})-(\ref{enyf2}), we obtain%
\begin{align}
\frac{\mathcal{E}^{2}-{}M^{2}}{2\eta _{2}}& =\left( 2n+1+\left\vert
m+s+s\eta _{1}\right\vert \right) +s\left( m+s\right) +\eta _{1}-1,
\label{enyg1} \\
\frac{\mathcal{E}^{2}-{}M^{2}}{2\eta _{2}}& =\left( 2n+1-\left\vert
m+s+s\eta _{1}\right\vert \right) +s\left( m+s\right) +\eta _{1}-1,
\label{enyg2}
\end{align}%
with unnormalized bound state wave functions given by%
\begin{eqnarray}
g_{m}(\tilde{\rho}) &=&\tilde{\rho}^{\frac{\left\vert \bar{\nu}\right\vert }{%
2}}e^{-\frac{\tilde{\rho}}{2}}F\left( -\bar{n},1+\left\vert \bar{\nu}%
\right\vert ,\tilde{\rho}\right)   \notag \\
&+&\tilde{\rho}^{-\frac{\left\vert \bar{\nu}\right\vert }{2}}e^{-\frac{%
\tilde{\rho}}{2}}F\left( -\bar{n},1-\left\vert \bar{\nu}\right\vert ,\tilde{%
\rho}\right) ,
\end{eqnarray}%
where%
\begin{equation}
\bar{n}=\frac{1\pm \left\vert \bar{\nu}\right\vert }{2}-\frac{\bar{E}}{4\eta
_{2}}.
\end{equation}%
If $\eta _{1}\rightarrow 0$ in Eqs. (\ref{enyf1})-(\ref{enyf2}) and (\ref%
{enyg1})-(\ref{enyg2}), we obtain
\begin{eqnarray}
\frac{\mathcal{E}^{2}-M^{2}}{2\eta _{2}} &=&\left( 2n+1+\left\vert
m\right\vert \right) +ms+1, \\
\frac{\mathcal{E}^{2}-{}M^{2}}{2\eta _{2}} &=&\left( 2n+1-\left\vert
m\right\vert \right) +ms+1,
\end{eqnarray}%
and%
\begin{align}
\frac{\mathcal{E}^{2}-{}M^{2}}{2\eta _{2}}& =\left( 2n+1+\left\vert
m+s\right\vert \right) +sm-1, \\
\frac{\mathcal{E}^{2}-{}M^{2}}{2\eta _{2}}& =\left( 2n+1-\left\vert
m+s\right\vert \right) +sm-1.
\end{align}%
These energy levels correspond to the analogue Landau quantization for
relativistic quantum dynamics of neutral fermions of spin-1/2 with magnetic
moment $\mathbf{\mu }$ in the field configuration of Eq. (\ref{fEb}).

\section{Nonrelativisitic limit}

\label{nonrel}

Let us now examine the nonrelativistic limit of Eq. (\ref{pauliequation}) by
setting $\mathcal{E}={}M+\varepsilon $, with $\varepsilon \ll {}M$. The
equation to be solved is
\begin{equation}
\frac{1}{2M}\left[ \mathbf{p}+s(\mathbf{\mu }\times \mathbf{E})\right]
^{2}\psi +\frac{1}{2M}\mu \sigma _{3}(\mathbf{\nabla }\cdot \mathbf{E})\psi
=\varepsilon \psi .  \label{pauliosnr}
\end{equation}%
Performing the same steps as for the relativistic case, we find the energy
levels%
\begin{equation}
\varepsilon =\left\{
\begin{array}{l}
\omega \left[ \left( 2n+1+\left\vert m+s\eta _{1}\right\vert \right) +\eta
_{1}+ms+1\right] , \\
\omega \left[ \left( 2n+1-\left\vert m+s\eta _{1}\right\vert \right) +\eta
_{1}+ms+1\right] .%
\end{array}%
\right.   \label{eigenup}
\end{equation}%
where we have defined the frequency $\omega =\eta _{2}/M$. Similarly, we can
also find the eigenvalues of Eq. (\ref{pauli2}). The result is given by%
\begin{equation}
\bar{\varepsilon}=\left\{
\begin{array}{l}
\omega \left[ \left( 2n+1+\left\vert m+s+s\eta _{1}\right\vert \right)
+s\left( m+s\right) +\eta _{1}-1\right] , \\
\omega \left[ \left( 2n+1-\left\vert m+s+s\eta _{1}\right\vert \right)
+s\left( m+s\right) +\eta _{1}-1\right] .%
\end{array}%
\right.   \label{eigendown}
\end{equation}%
If $\eta _{1}\rightarrow 0$, we obtain the energy levels corresponding to a
neutral fermion of spin-1/2 with magnetic moment in the nonrelativistic
regime
\begin{equation}
\varepsilon =\left\{
\begin{array}{l}
\omega \left[ \left( 2n+1+\left\vert m\right\vert \right) -ms+1\right] , \\
\omega \left[ \left( 2n+1-\left\vert m\right\vert \right) -ms+1\right] .%
\end{array}%
\right.   \label{nrla}
\end{equation}%
and%
\begin{equation}
\bar{\varepsilon}=\left\{
\begin{array}{l}
\omega \left[ \left( 2n+1+\left\vert m+s\right\vert \right) +s\left(
m+s\right) -1\right] , \\
\omega \left[ \left( 2n+1-\left\vert m+s\right\vert \right) +s\left(
m+s\right) -1\right] .%
\end{array}%
\right.   \label{nrlb}
\end{equation}%
Equations (\ref{nrla}) and (\ref{nrlb}) can be compared, for example, with
Eq. (25) of Ref. \cite{EPJC.2008.56.597}, in the absence of the spin element
$s$.

\section{Conclusions}

\label{conc}

We have solved the quantum dynamics of a neutral fermion with a magnetic
moment $\mu $ in the presence of external electric fields. We shown that the
set of first order differential equations admit isolated solutions ($E=M$,
and $E=-M$). This result implies new solutions for the AC problem. We derive
the second-order Dirac equation to study the motion of the particle in two
situations. First, we assume that the particle describes a circular path of
constant radius, and then analyze the dynamics in the full space, including
the $r=0$ region. The inclusion of the $r=0$ region states that we must
consider the term $\mathbf{\nabla }\cdot \mathbf{E}_{1}$ in Eq. (\ref%
{pauliequation}), i.e., a singular term. We consider the self-adjoint
extension method and show that the term $\mathbf{\nabla }\cdot \mathbf{E}_{1}
$, which results in a $\delta $ function, has physical implications on the
dynamics of the particle. In other words, we have verified that this term
contributes for the bound state wave function and energy spectrum, and with
an explicit dependence on the spin projection parameter $s$. For $\lambda
_{m}=0$, case in which the $\delta $ function is absent, only the regular
solution contributes for the bound state wave function. For two particular
values for the self-adjoint extension parameter, $\lambda _{m}=0$ (regular
solution) and $\lambda _{m}=\infty $ (irregular solution), the energies are
given explicitly in Eqs. (\ref{enyf1})-(\ref{enyf2}) and (\ref{enyg1})-(\ref%
{enyg2}). In the limit $\eta _{1}\rightarrow 0$, the corresponding energy
levels are analogous to Landau levels. In this limit, the dependence on s
parameter is still maintained. We also have obtained these results in
nonrelativistic limit.

\section*{Acknowledgments}

This work was supported by the CNPq, Brazil, Grants No. 482015/2013-6
(Universal), No 455719/2014-4 (Universal), No. 476267/2013-7 (Universal),
No. 306068/2013-3 (PQ), 304105/2014-7 (PQ) and FAPEMA, Brazil, Grants No.
00845/13 (Universal).

\bibliographystyle{spphys}
\bibliography{bibliography}

\end{document}